\documentclass{PoS}
\usepackage{multirow}

\title{Solar, supernova, atmospheric and geo neutrino studies using JUNO detector}

\ShortTitle{Solar, SN, atm, geo $\nu$ prospects at JUNO}

\author{\speaker{G.~Salamanna}\thanks{On behalf of the JUNO Collaboration}\\
        Roma Tre University and INFN Roma Tre, Rome, Italy \\
        E-mail: \email{salaman@cern.ch}}

\author{W.~L.~Guo\\
        IHEP, Beijing, China\\
}
\author{R.~Han\\
        BISEE, Beijing, China\\
}
\author{Y.~F.~Li\\
        IHEP, Beijing, China\\
}

\abstract{Aside from its primary purpose of shedding light on the mass hierarchy (MH) using
reactor anti-neutrinos, the JUNO experiment in Jiangmen (China) will also contribute to
study neutrinos from non-reactor sources.
In this poster we review JUNO's goals in the realms of supernova,
atmospheric, solar and geo-neutrinos; present the related experimental issues and
provide the current estimates of its potential.
For a typical galactic SN at a distance of 10 kpc, JUNO will record about 5000 events
from inverse beta decay, 2000 events from elastic neutrino-proton scattering, 300
events from neutrino-electron scattering, and the charged current and neutral current
interactions on the ${^{12}}{\rm C}$ nuclei. 
For atmospheric neutrinos, JUNO should be able to detect $\nu_e$ and
$\nu_\mu$ charged current events. 
Optimistically, a determination of the MH could be achieved at the 1.8$\sigma$
(2.6$\sigma$) level after 10 (20) years of data taking. JUNO will also study solar neutrinos from
${^{7}}{\rm Be}$ and ${^{8}}{\rm B}$, at low ($\approx$1 MeV) and higher energies respectively, to improve our
understanding of the matter effects on the oscillation mechanism and of the solar
metallicity. Challenges come primarily from the radioactive and cosmogenic backgrounds: the expected 
performance for two benchmark scintillator radio-purities, are shown. 
The flux of geo-neutrinos gives us an insight on the Earth composition and
formation. We will show how the increased sample size given by JUNO's large sensitive
mass of 20 KTon liquid scintillator will provide data to answer to several geological
questions among which the U/Th ratio and mantle measurements}

\FullConference{38th International Conference on High Energy Physics\\
		3-10 August 2016\\
		Chicago, USA}

\begin{document}

\section{Introduction}
JUNO is a neutrino experiment being built in China, described in \cite{An:2015jdp}. 
Its primary purpose is to determine the neutrino mass hierarchy (MH) and measure the oscillation parameters using reactor sources. In order to pursue these aims it is necessary to attain an unprecedented resolution on the energy of the $\bar{\nu}_e$ produced in the reactors: the goal is to achieve a resolution of $3\%$ at 1 MeV. These constraints render JUNO suitable also for the study of neutrinos originating from other sources: the Sun, the Earth centre and atmosphere and supernova explosions. Each of these topics bears relevant astrophysical or geophysical meaning. In the poster presented the potential of JUNO to detect and measure the main parameters connected with such sources was described.

\section{Atmospheric neutrinos}
\begin{figure}[htb]
\begin{center}
\includegraphics[scale=0.4]{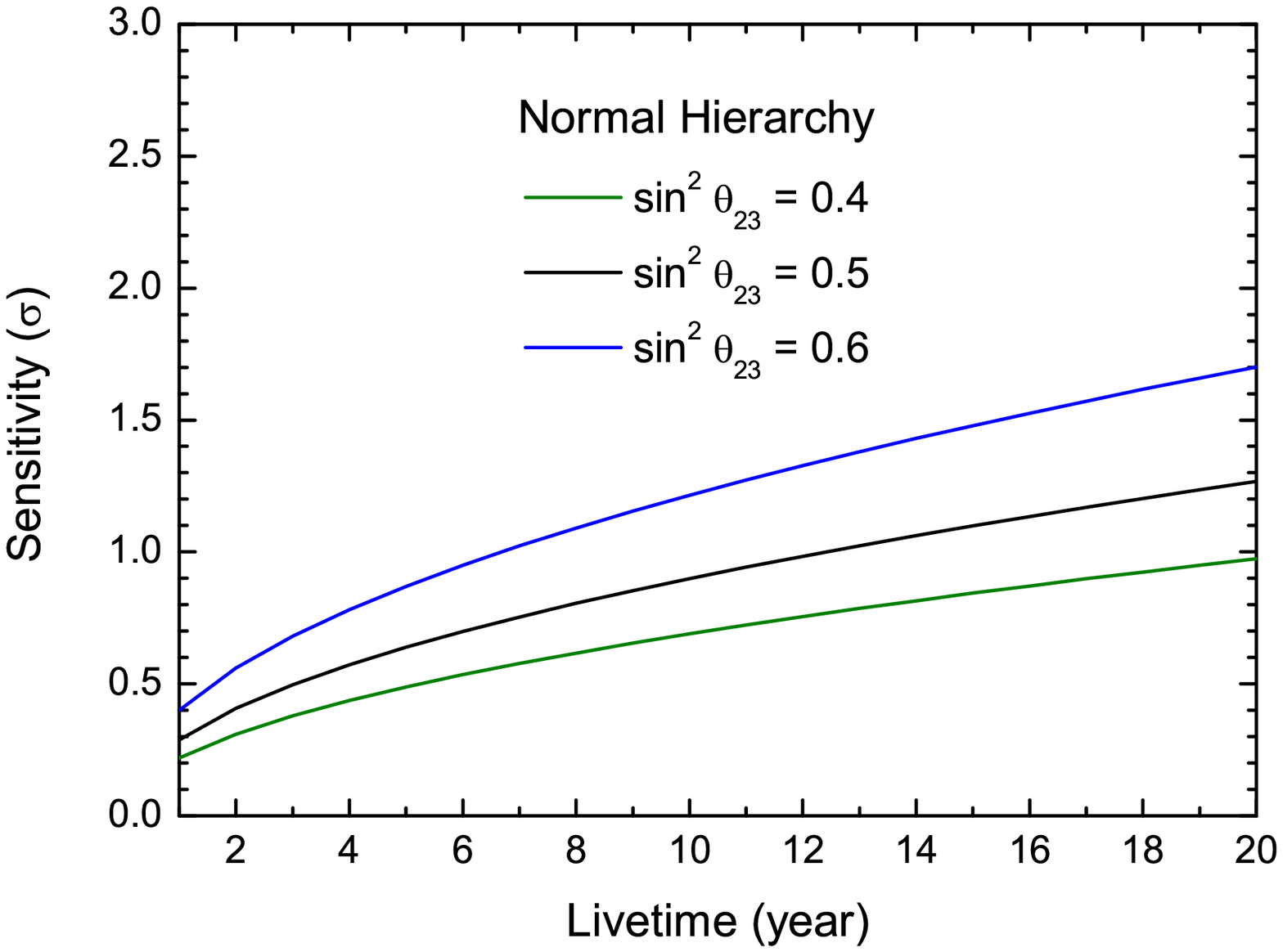}
\includegraphics[scale=0.4]{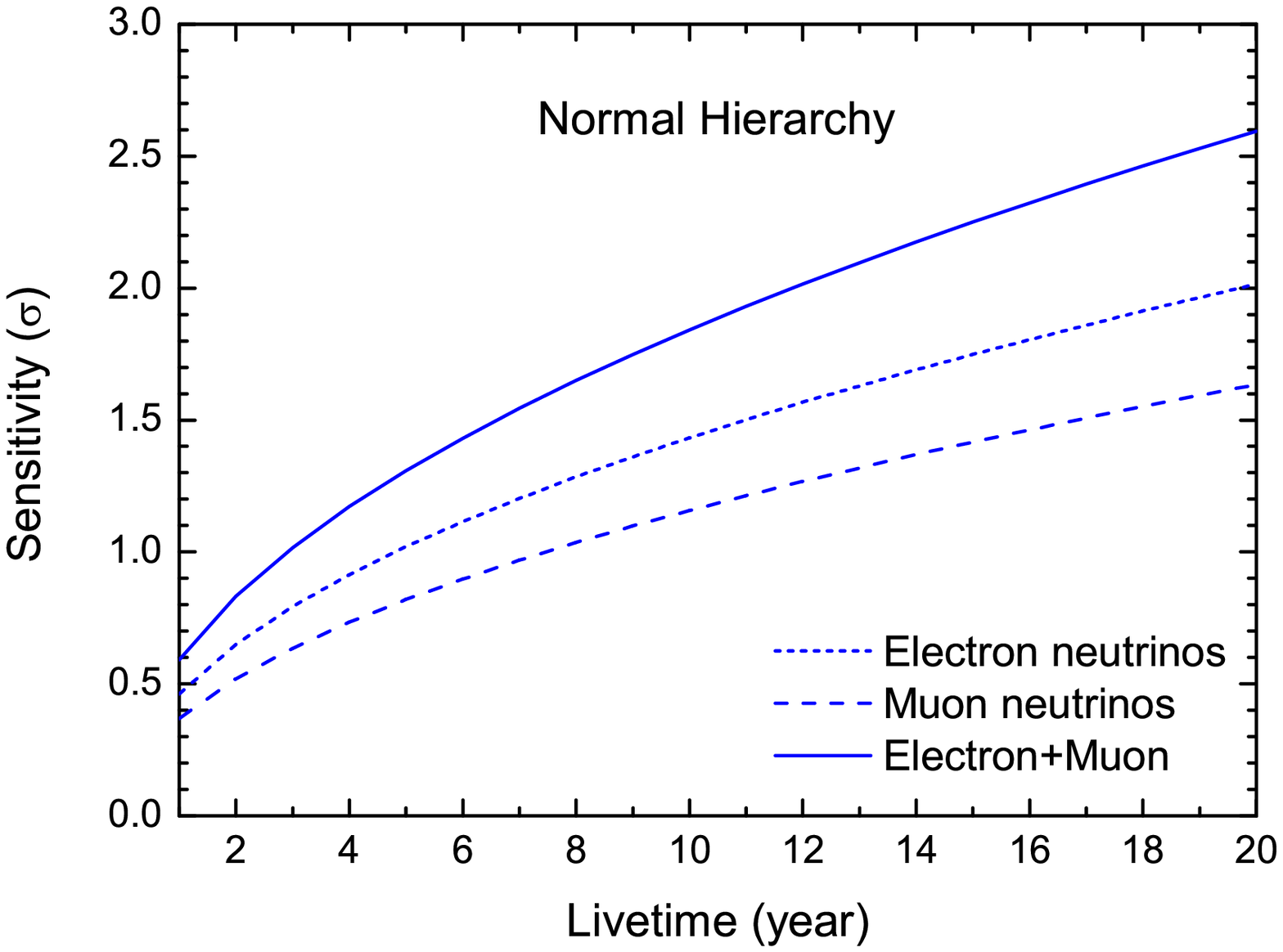}
\end{center}
\vspace{-0.5cm} \caption{ The JUNO conservative (left) and
optimistic (right) MH sensitivities as a function of livetime for
the true normal hierarchy hypothesis. } \label{results}
\end{figure}

The JUNO central detector has a very low energy threshold and can
measure atmospheric neutrinos with excellent energy resolution.
Characteristic signals from Michel electrons, neutron captures and
unstable daughter nuclei are helpful for the particle recognition.
The JUNO liquid scintillator (LS) detector has also some capabilities to reconstruct the
directions of charged leptons in terms of the timing pattern of the
first-hit on the photo-multiplier tubes (PMT) \cite{Learned}. 
Conservatively, only the atmospheric
$\nu_\mu/\bar{\nu}_\mu$ charged current (CC) events with a
$\mu^\pm$ track length $L_{\mu}
> 5 $ m are used to study the extraction of the MH from atmospheric neutrinos. 
For selected events, the assumption is that the final state $\mu^\pm$ can be fully
reconstructed and identified. The corresponding visible energy and
$\mu^\pm$ angular resolutions are taken to be $\sigma_{E_{vis}} =
0.01 \sqrt{E_{vis}/{\rm GeV}}$ and $1^\circ$, respectively.
Events are divided into four categories, according to the muon track length
and the $\nu/\bar{\nu}$ statistical separation \cite{Huber:2008yx}.
A $0.9 \sigma$ MH sensitivity can be achieved with 10 year data
and for $\sin^2 \theta_{23} = 0.5$, as shown in the left panel of Fig.
\ref{results}. The inverted hierarchy case has similar results
\cite{An:2015jdp}. An optimistic scenario for the neutrino MH is also contemplated.
Firstly, the $\nu_e/\bar{\nu}_e$ CC events can be identified and
reconstructed very well in the $e^\pm$ visible energy $E^e_{vis} >
1$ GeV and $E^e_{vis}/E_{vis} > 0.5$ case. Furthermore, a looser $L_{\mu} > 3 $m requirement 
is applied. Finally, the charged lepton direction is replaced with the
full neutrino direction, with a $10^\circ$ angular resolution. As
shown in right panel of Fig. \ref{results}, the combined sensitivity
can reach 1.8 $\sigma$ for 10 year data \cite{An:2015jdp}.

\section{Solar neutrinos}
\begin{figure}
  \centering
  \includegraphics[width=0.45\textwidth]{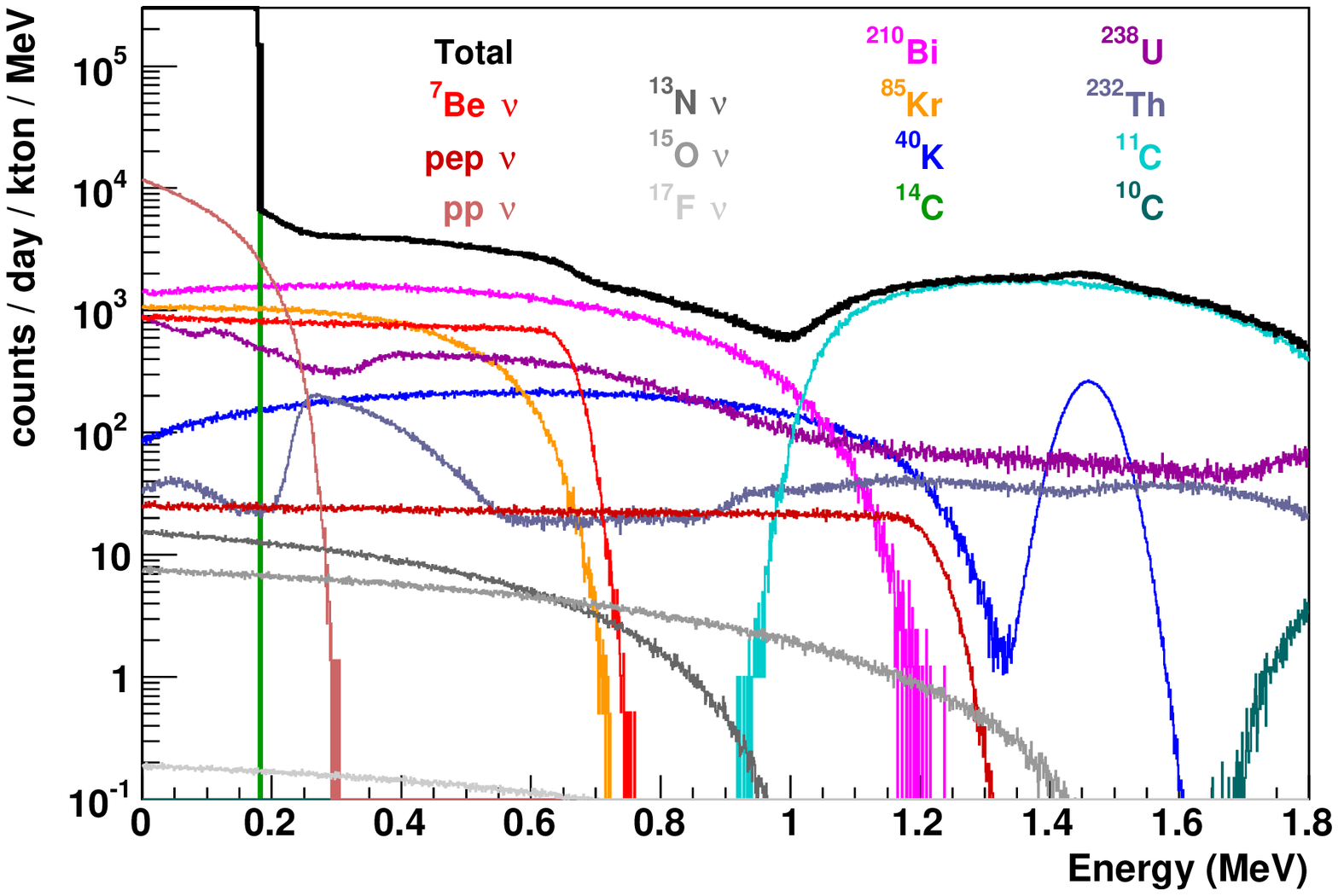}
  \includegraphics[width=0.45\textwidth]{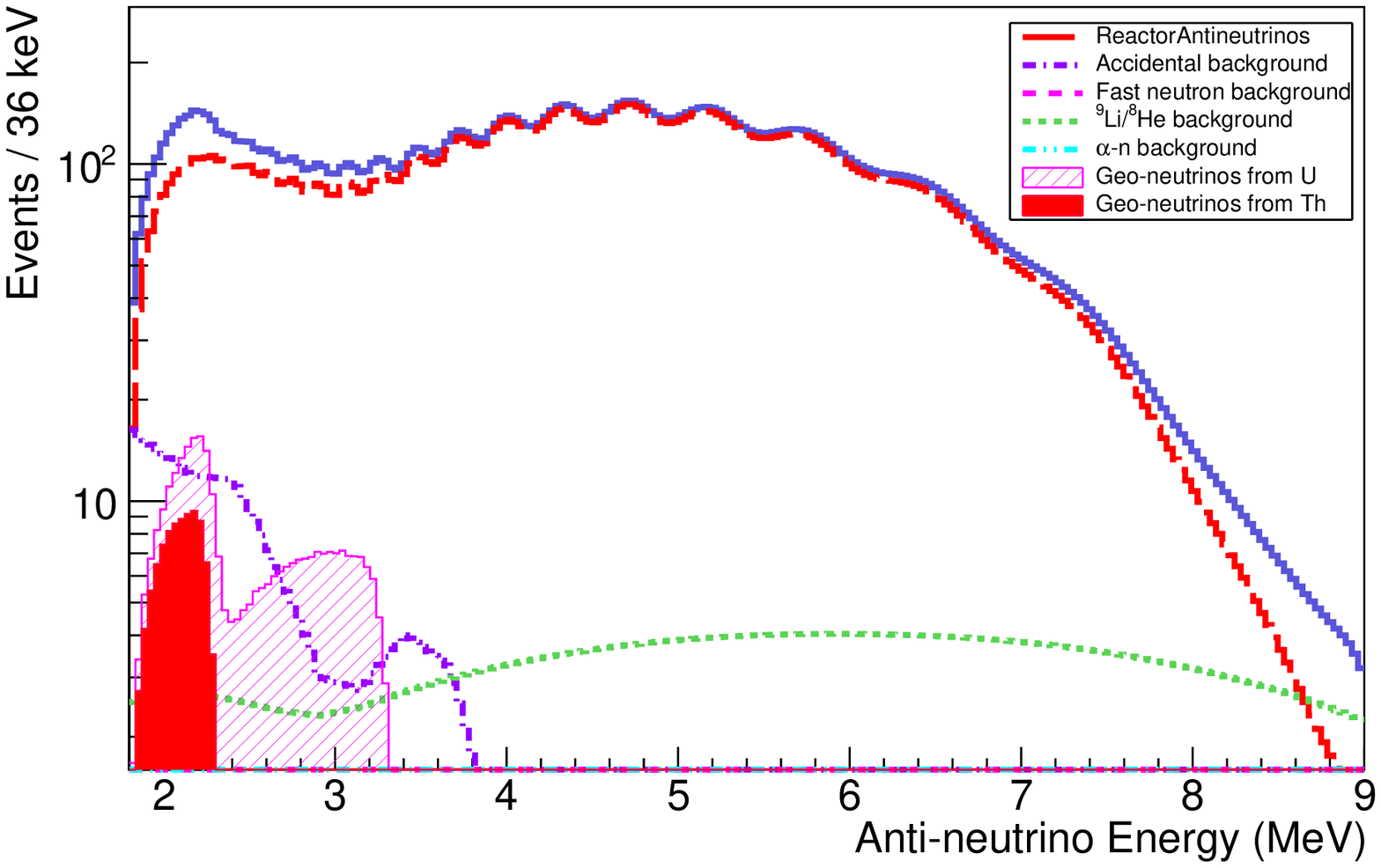}
  \caption{Left: Energy spectra of electrons resulting from solar neutrino elastic scattering and relevant backgrounds. The $\nu_e$ fluxes are taken from \cite{An:2015jdp}.
    Right: Energy spectra of geo-neutrinos, reactor antineutrinos, and other non-antineutrino backgrounds at JUNO for one year of data-taking. The blue solid line is the total spectrum, while the red dashed line is the reactor antineutrinos. The red solid area and pink area with parallel lines are antineutrinos from Th and U in the Earth, respectively.}
  \label{fig:solar_geo_spectra}
\end{figure}

The Sun is a powerful source of electron neutrinos with an energy of O(1) MeV, produced in two thermo-nuclear fusion reactions in the solar core (${\rm pp}$ chain and CNO cycle).
The relative importance of the two can be used to infer a star size: in the Sun, the ${\rm pp}$ chain constitutes around 99\% of the neutrino flux,
with the ${\rm pp}$, ${\rm pep}$, ${\rm hep}$, ${^{7}}{\rm Be}$ and ${^{8}}{\rm B}$ $\nu_e$ sources. The experimental signature is constituted by a single electron
scattered elastically by the incoming solar neutrino.
From studying the solar neutrino flux and energy spectrum important issues can be probed, such as testing of the Mikheyev-Smirnov-Wolfenstein (MSW) \cite{wolf},\cite{mikheev} matter effect in particle physics, particularly at energies sensitive to new physics contributions ($E_\nu\approx 2-5~{\rm MeV}$ \cite{maltoni}); and the agreement between solar models and the data from helioseismology \cite{villante}.
Thanks to its large exposure and the good energy resolution, JUNO can compete with the recent main actors in the sector (e.g. \cite{bx}). It can contribute to the measurement of $E_\nu$ from ${^{7}}{\rm Be}$ and discriminate the ${\rm pp}$ neutrinos from the sharply decreasing ${^{14}}{\rm C}$ spectrum. The main challenge is represented by the backgrounds: the intrinsic radioactivity of elements decaying in the LS, the acrylic vessel and the instrumentations; and from the cosmic ray muons interacting with ${^{12}}{\rm C}$ and producing lighter isotopes such as the long-lived ${^{11}}{\rm C}$ and ${^{10}}{\rm C}$. The latter are particularly relevant for ${\rm pep}$ and for ${^{8}}{\rm B}$ and CNO solar $\nu$ respectively. The assumed baseline (ideal) radio-purity, after LS purification, corresponds to a signal-to-background ratio S:B$\approx$1:3 (2:1). The resulting background and signal energy spectra are shown in Figure~\ref{fig:solar_geo_spectra} for the baseline case.

\section{Supernova neutrinos}
A massive star of mass above eight solar masses is expected to experience a core collapse under its own gravity, and then a violent explosion,
where 99\% of the gravitational binding energy of the core-collapse supernova (SN) will be carried by the intense burst of neutrinos~\cite{Janka:2012wk}.
Measuring the neutrino burst from the next nearby SN is a primary goal for the studies of both neutrino physics and astrophysics.
As the largest LS detector of new generation, JUNO will be superior in its high statistics, the good energy
resolution and rich neutrino flavor information~\cite{An:2015jdp}. For a Galactic SN burst at 10 kpc,
JUNO will register about 5000 events from inverse beta decay (IBD),
$\overline{\nu}^{}_e + p \to n + e^+$, 1000 events from all-flavor elastic neutrino-proton scattering, $\nu + p \to \nu + p$,
more than 300 events from neutrino-electron scattering, $\nu + e^- \to \nu + e^-$, as well as the charged current and neutral current
interactions on the ${^{12}}{\rm C}$ nuclei.
The numbers of neutrino events at JUNO~\cite{An:2015jdp} for a SN at a typical distance of 10 kpc
are presented in Table~\ref{tab:events}.
With these measurements of SN neutrinos, JUNO may provide precious information to measure the initial SN neutrino fluxes~\cite{Lu:2016ipr},
to constrain the neutrino mass scale~\cite{Lu:2014zma} and ordering, to test the scenario of collective neutrino oscillations and even to probe the
neutrino electromagnetic properties~\cite{Giunti:2015gga}.

\begin{table}[!t]
\centering
\begin{small}
\begin{tabular}{ccccccccc}
\hline
\multicolumn{1}{c}{\multirow {2}{*}{Channel}} & \multicolumn{1}{c}{} & \multicolumn{1}{c}{\multirow {2}{*}{Type}} & \multicolumn{1}{c}{} & \multicolumn{5}{c}{Events for different $\langle E_\nu \rangle$ values} \\
\cline{5-9} \multicolumn{1}{c}{} & \multicolumn{1}{c}{} & \multicolumn{1}{c}{} & \multicolumn{1}{c}{} & \multicolumn{1}{c}{$12~{\rm MeV}$} & \multicolumn{1}{c}{} & \multicolumn{1}{c}{$14~{\rm MeV}$} & \multicolumn{1}{c}{} & \multicolumn{1}{c}{$16~{\rm MeV}$} \\
\hline
\multicolumn{1}{l}{$\overline{\nu}_e + p \to e^+ + n$} & \multicolumn{1}{c}{} & \multicolumn{1}{c}{CC} & \multicolumn{1}{c}{} & \multicolumn{1}{c}{$4.3\times 10^3$} & \multicolumn{1}{c}{} & \multicolumn{1}{c}{$5.0\times 10^3$} & \multicolumn{1}{c}{} & \multicolumn{1}{c}{$5.7\times 10^3$} \\
\multicolumn{1}{l}{$\nu + p \to \nu + p$} & \multicolumn{1}{c}{} & \multicolumn{1}{c}{NC} & \multicolumn{1}{c}{} & \multicolumn{1}{c}{$0.6\times 10^3$} & \multicolumn{1}{c}{} & \multicolumn{1}{c}{$1.2\times 10^3$} & \multicolumn{1}{c}{} & \multicolumn{1}{c}{$2.0\times 10^3$} \\
\multicolumn{1}{l}{$\nu + e \to \nu + e$} & \multicolumn{1}{c}{} & \multicolumn{1}{c}{ES} & \multicolumn{1}{c}{} & \multicolumn{1}{c}{$3.6\times 10^2$} & \multicolumn{1}{c}{} & \multicolumn{1}{c}{$3.6\times 10^2$} & \multicolumn{1}{c}{} & \multicolumn{1}{c}{$3.6\times 10^2$} \\
\multicolumn{1}{l}{$\nu +~^{12}{\rm C} \to \nu +~^{12}{\rm C}^*$} & \multicolumn{1}{c}{} & \multicolumn{1}{c}{NC} & \multicolumn{1}{c}{} & \multicolumn{1}{c}{$1.7\times 10^2$} & \multicolumn{1}{c}{} & \multicolumn{1}{c}{$3.2\times 10^2$} & \multicolumn{1}{c}{} & \multicolumn{1}{c}{$5.2\times 10^2$} \\
\multicolumn{1}{l}{$\nu_e +~^{12}{\rm C} \to e^- +~^{12}{\rm N}$} & \multicolumn{1}{c}{} & \multicolumn{1}{c}{CC} & \multicolumn{1}{c}{} & \multicolumn{1}{c}{$0.5\times 10^2$} & \multicolumn{1}{c}{} & \multicolumn{1}{c}{$0.9\times 10^2$} & \multicolumn{1}{c}{} & \multicolumn{1}{c}{$1.6\times 10^2$} \\
\multicolumn{1}{l}{$\overline{\nu}_e +~^{12}{\rm C} \to e^+ +~^{12}{\rm B}$} & \multicolumn{1}{c}{} & \multicolumn{1}{c}{CC} & \multicolumn{1}{c}{} & \multicolumn{1}{c}{$0.6\times 10^2$} & \multicolumn{1}{c}{} & \multicolumn{1}{c}{$1.1\times 10^2$} & \multicolumn{1}{c}{} & \multicolumn{1}{c}{$1.6\times 10^2$} \\
\hline
\end{tabular}
\end{small}
\caption{Numbers of neutrino events in JUNO for a SN at a typical distance of 10 kpc.
Three representative values of the average neutrino energy $\langle E_\nu \rangle = 12~{\rm MeV}$, $14~{\rm MeV}$ and $16~{\rm MeV}$ are taken for illustration.
For the elastic neutrino-proton scattering, a threshold of $0.2~{\rm MeV}$ for the proton recoil energy is chosen.}
\label{tab:events}
\end{table}


\section{Geo-neutrinos}
Geo-neutrinos are electron antineutrinos emitted from nuclei undergoing $\beta$ decay inside the Earth. Those from Th and U have the highest energies and undergo an inverse beta decay (IBD) in the JUNO LS. The flux of geo-neutrinos at any point on the Earth is a function of the abundance and distribution of radioactive elements within our planet. This flux has been successfully detected by the 1-kt KamLAND and 0.3-kt Borexino detectors, with these measurements being limited by their low statistics. With a much higher mass, JUNO should gather enough data to separate the contributions of Th and U. Its Th/U ratio can bring resolution to several major issues in Earth sciences: which are the sources of the power that is driving plate tectonics, mantle convection, and the geodynamo, as well the structure of mantle convection. The main background coming to geo-neutrino studies comes from nearby nuclear power plants. The expected energy spectra for geo-neutrino signal as well as reactor antineutrino and non-antineutrino backgrounds are shown in Figure~\ref{fig:solar_geo_spectra}. Scenarios where the Th/U ratio is fixed to the chondritic value or is free Th/U were considered. With 1, 3, 5, and 10 years of data, the precision of the geo-neutrino measurement with a fixed chondritic Th/U ratio is 13\%, 8\%, 6\% and 5\%, respectively, which as expected, decreases with higher statistics. For the future, we need to predict the geo-neutrino flux at the JUNO experimental site. The combined geological, geophysical and geochemical data will be integrated into a 3-dimensional reference model of the lithosphere~\cite{Huang:2013}. After building this geological model geologists and particle physicists will work together to calculate the regional geo-neutrino flux.

\section{Conclusions}
The poster presented illustrates the foreseen potential of the JUNO experiment for the measurement of properties of neutrinos from the sun, the Earth and astrophysical events. The main background sources and selection strategy have been described. Thanks to its very large exposure and expected unprecedented energy resolution, JUNO should be competitive and improve on past measurements, with a unique potential for e.g. supernova neutrinos.


\begin{thebibliography}{99}

\bibitem{An:2015jdp}
  F.~An {\it et al.} [JUNO Collaboration],
  J.\ Phys.\ G {\bf 43}, 030401 (2016)

\bibitem{Learned}
  J.~G.~Learned
  [arXiv:0902.4009]

\bibitem{Huber:2008yx}
  P.~Huber and T.~Schwetz
  Phys.\ Lett.\ B {\bf 669}, 294 (2008)

\bibitem{wolf}
  L.~Wolfenstein
  Phys.\ Rev.\ D {\bf 17}, 2369 (1978)

\bibitem{mikheev}
  S.~P.~Mikheyev, A.~Y.~Smirnov,
  Sov.\ J.\ Nucl.\ Phys.\ {\bf 42}, 913 (1985)

\bibitem{maltoni}
  M.~Maltoni, A.~Y.~Smirnov,
  Eur.\ Phys.\ J.\ A {\bf 52}, 87 (2016)

\bibitem{villante}
 F.~L.~Villante, A.~Serenelli, F.~Delahaye, M.~H.~Pinsonneault,
 Astrophys.\ J.\ {\bf 787}, 13 (2014)

\bibitem{bx}
 G.~Bellini {\it et al.} [Borexino Collaboration],
 Nature {\bf 512}, 383 (2014)

\bibitem{Janka:2012wk}
  H.~T.~Janka,
  Ann.\ Rev.\ Nucl.\ Part.\ Sci.\  {\bf 62}, 407 (2012)

\bibitem{Lu:2016ipr}
  J.~S.~Lu, Y.~F.~Li and S.~Zhou,
  Phys.\ Rev.\ D {\bf 94}, no. 2, 023006 (2016)

\bibitem{Lu:2014zma}
  J.~S.~Lu, J.~Cao, Y.~F.~Li and S.~Zhou,
  JCAP {\bf 1505}, no. 05, 044 (2015)

\bibitem{Giunti:2015gga}
  C.~Giunti {\it et al},
  Annalen Phys.\  {\bf 528}, 198 (2016)


\bibitem{Huang:2013}

Y.~Huang {\it et al},
 Geochemistry, Geophysics, Geosystems.\ {\bf 14 }, 2003¨C2029(2013)

\end{thebibliography}
\end{document}